\documentclass
[aps,floats,twoside,fancyhdr,a4paper,superscriptaddress,nofootinbib,showpacs]{revtex4}%
\usepackage{latexsym}
\usepackage{amsmath}
\usepackage{amssymb}
\usepackage{amsfonts}
\usepackage{graphicx}
\usepackage{dcolumn}
\usepackage{fancyhdr}
\usepackage[colorlinks]{hyperref}%
\begin{document}
\title{Perfect Fluid FRW with Time-varying Constants Revisited.}
\author{J. A. Belinch\'on}
\email{abelcal@ciccp.es}
\affiliation{Dpt. of Physics, ETS Architecture UPM. Av. Juan de Herrera N-4. Madrid 28040 Espa\~{n}a.}
\author{J. L. Caram\'{e}s}
\email{JoseL\_Carames@mat.ucm.es}
\affiliation{Dpt. An\'alisis Matem\'atico, Facultad CC. Matem\'aticas UCM. Madrid 28040 Espa\~na.}
\date{\today }

\begin{abstract}
In this paper we revise a perfect fluid FRW model with time-varying constants
\textquotedblleft but\textquotedblright\ taking into account the effects of a
\textquotedblleft$c$-variable\textquotedblright\ into the curvature tensor. We
study the model under the following assumptions, $div(T)=0$ and $div(T)\neq0,$
and in each case the flat and the non-flat cases are studied. Once we have
outlined the new field equations, it is showed in the flat case i.e. $K=0$,
that there is a non-trivial homothetic vector field i.e. that this case is
self-similar. In this way, we find that there is only one symmetry, the
scaling one, which induces the same solution that the obtained one in our
previous model. At the same time we find that \textquotedblleft constants" $G$
and $c$ must verify, as integration condition of the field equations, the
relationship $G/c^{2}=const.$ in spite of that both \textquotedblleft
constants" vary. We also find that there is a narrow relationship between the
equation of state and the behavior of the time functions $G,c$ and the sign of
$\Lambda$ in such a way that these functions may be growing as well as
decreasing functions on time $t,$ while $\Lambda$ may be a positive or
negative decreasing function on time $t.$ In the non-flat case it will be
showed that there is not any symmetry. For the case $div(T)\neq0,$ it will be
studied again the flat and the non-flat cases. In order to solve this case it
is necessary to make some assumptions on the behavior of the time functions
$G,c$ and $\Lambda.$ We also find the flat case with $div(T)=0,$ is a
particular solution of the general case $div(T)\neq0.$

\end{abstract}

\maketitle

\section{Introduction}

Since the pioneering work of Dirac (\cite{1}), who proposed, motivated by the
occurrence of large numbers in Universe, a theory with a time variable
gravitational coupling constant $G$, cosmological models with variable $G$ and
nonvanishing and variable cosmological term have been intensively investigated
in the physical literature (see for example \cite{2}-\cite{14}).

Recently , the cosmological implications of a variable speed of light during
the early evolution of the Universe have been considered. Varying speed of
light (VSL) models proposed by Moffat (\cite{moff}) and Albrecht and Magueijo
(\cite{Magueijo0}), in which light was travelling faster in the early periods
of the existence of the Universe, might solve the same problems as inflation.
Hence they could become a valuable alternative explanation of the dynamics and
evolution of our Universe and provide an explanation for the problem of the
variation of the physical \textquotedblleft constants". Einstein's field
equations (EFE) for Friedmann--Robertson--Walker (FRW) spacetime in the VSL
theory have been solved by Barrow (\cite{Barrow}), who also obtained the rate
of variation of the speed of light required to solve the flatness and
cosmological constant problems (see J. Magueijo (\cite{Magueijo1}) for a
review of these theories).

This model is formulated under the strong assumption that a $c$ variable
(where $c$ stands for the speed of light) does not introduce any corrections
into the curvature tensor, furthermore, such formulation does not verify the
covariance and the Lorentz invariance as well as the resulting field equations
do not verify the Bianchi identities either (see Bassett et al \cite{Bassett}).

Nevertheless, some authors (T. Harko and M. K. Mak \cite{Harko1} and P.P.
Avelino and C.J.A.P. Martins \cite{Avelino}) have proposed a new
generalization of General Relativity which also allows arbitrary changes in
the speed of light, $c$, and the gravitational constant, $G$, but in such a
way that variations in the speed of light introduce corrections to the
curvature tensor in the Einstein equations in the cosmological frame. This new
formulation is both covariant and Lorentz invariant and as we will show the
resulting field equations (FE) verify the Bianchi identities. We will be able
to obtain the energy conservation equation from the field equations as in the
standard case.

The purpose of this paper is to revise a FRW perfect fluid model with time
varying constants (see \cite{to1} and \cite{to2}) but taking into account the
effects of a $c$ variable into the field equations. In our previous models
(\cite{to1}-\cite{to2}) we worked under the assumption that a $c-$variable
does not induce corrections into the curvature tensor and hence the classical
Friedman equations remain valid. We will show that such effect is minimum but
exists. In section 2, once we have outlined the Einstein tensor, we check that
it verifies that its covariant divergence vanishes. Consequently we impose the
same condition to the right hand of the Einstein equation i.e. $div\left(
\frac{8\pi G}{c^{4}}T^{ij}+g^{ij}\Lambda\right)  =0. $ In this way we set the
new field equations as well as the Kretschmann scalars. As we will see, in
order to integrate the resulting field equations we will need to consider some
assumptions. In section 3, we will consider the condition $div\left(
T^{ij}\right)  =0,$ and it will be studied two particular cases, the flat
case, i.e. $K=0$, and the non-flat case, $K\neq0. $ Since with the proposed
method we are not able to solve the so called flat problem we need to study
separately both cases (the flat and the non-flat cases) introducing such
conditions as an assumption.

In this approach, in the case $K=0,$ we will show that the model is
self-similar since it is found a non-trivial homothetic vector field. In order
to integrate the FE under the assumptions $div(T)=0$ and $K=0,$ we use the Lie
group tactic which allows us to find a particular form of $G$ and $c$ for
which our FE admit symmetries i.e. are integrable. As we will show under these
assumptions the FE, only admit one symmetry, the scaling symmetry. This is the
main difference with respect to our previous approach (\cite{to1}-\cite{to2})
where we did not consider the effects of a $c$-variable into the curvature
tensor in such a way that the resulting FE admitted more symmetries. We also
obtain as integration condition that the \textquotedblleft
constants\textquotedblright\ $G$ and $c$ must verify the relationship
$G/c^{2}=const.$ in spite of that both \textquotedblleft
constants\textquotedblright\ vary. In this work we have found three solutions.
The first one is very similar to the de-Sitter solution i.e. the energy
density vanishes, the cosmological constant is a true constant while $G$ and
$c$ follow an exponential law as the scale factor. The second and third
solutions behave as the standard FRW where all the quantities follow power law
with respect to time $t,$ nevertheless the third solution is non-singular. In
the second studied case, the non-flat case $K\neq0,$ we will show that the FE
do not admit any symmetry. Nevertheless we will try to find a particular
solution imposing some restrictions.

In section 4 we consider the possibility that $div(T)\neq0,$ and we will study
again two particular cases, the flat case, i.e. $K=0$, and the non-flat case,
$K\neq0.$ The possibility that the covariant conservation condition $div(T)=0$
be relaxed has been advanced by Rastall (\cite{Rastall}), who pointed out that
a non-zero divergence of the energy--momentum tensor has not been ruled out
experimentally at all yet. In order to integrate the resulting FE we will need
to make some assumptions (scaling assumptions) about the behavior of the
\textquotedblleft constants\textquotedblright\ $G,c $ and $\Lambda,$ which
allow us to find particular solutions to the FE. This scaling assumptions work
well in the flat (homothetic) case, as it is expected, but they seem very
restrictive in the non-flat case. It is also showed that we can recover the
solution obtained in the $div(T)=0$ as a particular solution of this model in
such a way that $G$ and $c$ must verify the relationship $G/c^{2}=const.$
etc... as it is expected.

In section 5 we end with a brief conclusions.

\section{The model}

We will use the field equations in the form:
\begin{equation}
R_{ij}-\frac{1}{2}g_{ij}R=\frac{8\pi G(t)}{c^{4}(t)}T_{ij}+\Lambda
(t)g_{ij},\label{ECU1}%
\end{equation}
where arbitrary variations in $c$ and $G$ will be allowed. We assume that
variations in the speed of light introduce corrections to the curvature terms
in the Einstein equations in the cosmological frame. In our model variations
in the velocity of light are always allowed to contribute to the curvature
terms. These contributions are computed from the metric tensor in the usual
way. The line element is defined by (we are following the O'Neill's notation
\cite{o'neil}):
\begin{equation}
ds^{2}=-c(t)^{2}dt^{2}+f^{2}(t)\left[  \frac{dr^{2}}{1-Kr^{2}}+r^{2}\left(
d\theta^{2}+\sin^{2}\theta d\phi^{2}\right)  \right]  ,
\end{equation}
and the energy momentum tensor is:
\begin{equation}
T_{ij}=\left(  \rho+p\right)  u_{i}u_{j}-pg_{ij},\label{s2-e2}%
\end{equation}
where $p$ and $\rho$ satisfie the usual equation of state, $p=\omega\rho$ in
such a way that $\omega=const.$, usually $\omega$ is taken such that,
$\omega\in\left(  -1,1\right]  ,$ that is to say, our universe is modeled by a
perfect fluid$.$ The $4-$velocity $u^{i}$ is defined as follows $u^{i}=\left(
c^{-1}(t),0,0,0\right)  $ such that $u_{i}u^{i}=-1.$

The field equations are as follows:
\begin{align}
2\frac{f^{\prime\prime}}{f}-2\frac{c^{\prime}}{c}\frac{f^{\prime}}{f}%
+\frac{f^{\prime2}}{f^{2}}+\frac{Kc^{2}}{f^{2}}  & =-\frac{8\pi G}{c^{2}%
}p+\Lambda c^{2},\label{naomi}\\
\frac{f^{\prime2}}{f^{2}}+\frac{Kc^{2}}{f^{2}}  & =\frac{8\pi G}{3c^{2}}%
\rho+\frac{1}{3}\Lambda c^{2},
\end{align}
where as we can see, the only difference with respect to the usual one is the
factor $-2\frac{c^{\prime}}{c}\frac{f^{\prime}}{f}$ in eq. (\ref{naomi}). Here
a prime denotes differentiation with respect to time $t.$

Since the divergence of $\left(  R_{ij}-\frac{1}{2}g_{ij}R\right)  $ vanishes
then we impose that the right hand of equation (\ref{ECU1}) has zero
divergence too. Therefore applying the covariant divergence to the second
member of the field equation we get:
\begin{equation}
div\left(  \frac{8\pi G}{c^{4}}T^{ij}+g^{ij}\Lambda\right)  =0,
\end{equation}
where we are considering that \ $\frac{8\pi G}{c^{4}}$ is a function on time
$t$. Hence simplifying it yields:
\begin{equation}
\rho^{\prime}+3\left(  \omega+1\right)  \rho H=-\left(  \frac{G^{\prime}}%
{G}-4\frac{c^{\prime}}{c}\right)  \rho-\frac{\Lambda^{\prime}c^{4}}{8\pi
G},\label{DEF3}%
\end{equation}
where $H=f^{\prime}/f.$

Therefore the new field equations are as follows:
\begin{align}
2H^{\prime}-2\frac{c^{\prime}}{c}H+3H^{2}+\frac{Kc^{2}}{f^{2}}  & =-\frac{8\pi
G}{c^{2}}p+\Lambda c^{2},\label{nf1}\\
H^{2}+\frac{Kc^{2}}{f^{2}}  & =\frac{8\pi G}{3c^{2}}\rho+\frac{1}{3}\Lambda
c^{2},\label{nf2}\\
\rho^{\prime}+3\left(  \omega+1\right)  \rho H  & =-\left(  \frac{G^{\prime}%
}{G}-4\frac{c^{\prime}}{c}\right)  \rho-\frac{\Lambda^{\prime}c^{4}}{8\pi
G},\label{def_nf3}%
\end{align}

As we have mentioned in the introduction these field equations are not new in
the literature. They have been outlined by T. Harko and M. Mak (\cite{Harko1})
where they study a perfect fluid model and the influence of the time variation
of the constants in the matter creation. Later they study Bianchi I and V
model with time varying constants (taking into account the influence of a
c-variable into the curvature tensor). Other author are P.P. Avelino and
C.J.A.P. Martins (\cite{Avelino}) and\ H. Shojaie and M. Farhoudi, have
obtained similar equations (\cite{Iranies}) since these authors consider $G$
as a true constant.

We would like to emphasize that deriving eq. (\ref{nf2}) and substituting this
result into eq. (\ref{nf1}) it is obtained eq.(\ref{DEF3}) i.e. the covariant
divergence of the right hand of our field equation in the same way as in the
standard cosmological model.

Curvature is described by the tensor field $R_{jkl}^{i}.$ It is well known
that if one uses the singular behavior of the tensor components \ or its
derivates as a criterion for singularities, one gets into trouble since the
singular behavior of the coordinates or the tetrad basis rather than the
curvature tensor. In order to avoid this problem, one should examine the
scalars formed out of the curvature. The invariants $RiemS$ and $RiccS$ (the
Kretschmann scalars) are very useful for the study of the singular behavior,
being these as follows:%
\begin{equation}
RiemS=R_{ijlm}R^{ijlm},
\end{equation}%
\begin{equation}
RiemS:=\dfrac{12}{c^{4}}\left[  \left(  \frac{f^{\prime\prime}}{f}\right)
^{2}-2\frac{f^{\prime\prime}}{f}\frac{f^{\prime}}{f}\frac{c^{\prime}}%
{c}+\left(  \frac{f^{\prime}}{f}\right)  ^{2}\left(  \frac{c^{\prime}}%
{c}\right)  ^{2}+\left(  \frac{f^{\prime}}{f}\right)  ^{4}+2\left(
\frac{f^{\prime}}{f}\right)  ^{2}\dfrac{c^{2}K}{f^{2}}+\dfrac{c^{4}K}{f^{4}%
}\right]  ,
\end{equation}
and
\[
RiccS:=R_{ij}R^{ij},
\]%
\begin{align}
RiccS  & :=\dfrac{12}{c^{4}}\left[  \left(  \frac{f^{\prime\prime}}{f}\right)
^{2}-2\frac{f^{\prime\prime}}{f}\frac{f^{\prime}}{f}\frac{c^{\prime}}%
{c}+\left(  \frac{f^{\prime}}{f}\right)  ^{2}\left(  \frac{c^{\prime}}%
{c}\right)  ^{2}+\left(  \frac{f^{\prime}}{f}\right)  ^{4}+2\left(
\frac{f^{\prime}}{f}\right)  ^{2}\dfrac{c^{2}K}{f^{2}}+\dfrac{c^{4}K}{f^{4}%
}\right. \nonumber\\
& \left.  \frac{f^{\prime\prime}}{f}\left(  \dfrac{c^{2}K}{f^{2}}+\left(
\frac{f^{\prime}}{f}\right)  ^{2}\right)  -\frac{f^{\prime}}{f}\frac
{c^{\prime}}{c}\dfrac{c^{2}K}{f^{2}}-\frac{c^{\prime}}{c}\left(
\frac{f^{\prime}}{f}\right)  ^{3}\right]  .
\end{align}

In order to try to solve eqs. (\ref{nf1})-(\ref{def_nf3}) we need to make some
hypotheses on the behavior of the quantities.

\section{Solution I. $divT=0.$}

In this first solution we make the following assumption
\begin{equation}
divT=0
\end{equation}
and we will consider two subclasses, the flat and the non-flat cases.

Our tactic consists in studying the field equations through the Lie method
(see \cite{Stephani et al.}$-$\cite{Ibra}$-$\cite{Blu}). In particular we seek
the forms of $G$ and $c$ for which our field equations admit symmetries i.e.
are integrable.

We will consider the following assumption $divT=0,$ which transforms
eq.(\ref{def_nf3}) into these two new equations:%
\begin{align}
\rho^{\prime}+3\left(  \omega+1\right)  \rho H  & =0,\label{nf3}\\
\frac{G^{\prime}}{G}-4\frac{c^{\prime}}{c}  & =-\frac{\Lambda^{\prime}c^{4}%
}{8\pi G\rho}.\label{nf4}%
\end{align}

In order to use the Lie method, we rewrite the field equations as follows.
From (\ref{nf1} and \ref{nf2}) we obtain
\begin{equation}
2H^{\prime}-2\frac{c^{\prime}}{c}H-2\frac{Kc^{2}}{f^{2}}=-\frac{8\pi G}{c^{2}%
}\left(  p+\rho\right)  ,
\end{equation}
From equation (\ref{nf3}), we can obtain
\begin{equation}
H=-\frac{1}{3\left(  \omega+1\right)  }\frac{\rho^{\prime}}{\rho
}\Longrightarrow H=-C_{1}\frac{\rho^{\prime}}{\rho},
\end{equation}
where $C_{1}=\frac{1}{3\left(  \omega+1\right)  },$ therefore
\begin{equation}
-2C_{1}\left(  \frac{\rho^{\prime\prime}}{\rho}-\frac{\rho^{\prime2}}{\rho
^{2}}\right)  +2C_{1}\frac{\rho^{\prime}}{\rho}\frac{c^{\prime}}{c}%
-2\frac{Kc^{2}}{f^{2}}=-\frac{8\pi G}{c^{2}}\left(  \omega+1\right)  \rho,
\end{equation}
hence
\begin{equation}
\rho^{\prime\prime}=\frac{\rho^{\prime2}}{\rho}+\rho^{\prime}\frac{c^{\prime}%
}{c}-2\frac{Kc^{2}}{f^{2}}\rho+A\frac{G}{c^{2}}\rho^{2},\label{Lie-general}%
\end{equation}
where $A=12\pi\left(  \omega+1\right)  ^{2}>0,\forall\omega.$ Following this
tactic we try to make the smallest hypothesis number and to obtain the exact
behavior of the \textquotedblleft constants\textquotedblright\ $G,c$ and
$\Lambda.$Following other tactics we are obliged to make assumptions that
could be unphysical.

\subsection{Flat case. Self-similar approach.}

As have been pointed out by Carr and Coley (\cite{20}), the existence of self-
similar solutions (Barenblatt and Zeldovich (\cite{21})) is related to
conservation laws and to the invariance of the problem with respect to the
group of similarity transformations of quantities with independent dimensions.
This can be characterized within general relativity by the existence of a
homothetic vector field and for this reason one must distinguish between
geometrical and physical self-similarity. Geometrical similarity is a property
of the spacetime metric, whereas physical similarity is a property of the
matter fields (our case). In the case of perfect fluid solutions admitting a
homothetic vector, geometrical self-similarity implies physical self-similarity.

As we show in this section as well as in previous works, the assumption of
self-similarity reduces the mathematical complexity of the governing
differential equations. This makes such solutions easier to study
mathematically. Indeed self-similarity in the broadest Lie sense refers to an
invariance which allows such a reduction.

Perfect fluid space-times admitting a homothetic vector within general
relativity have been studied by Eardley (\cite{22}). In such space-times, all
physical transformations occur according to their respective dimensions, in
such a way that geometric and physical self-similarity coincide. It is said
that these space-times admit a transitive similarity group and space-times
admitting a non-trivial similarity group are called self-similar. Our model
i.e. a flat FRW model with a perfect fluid stress-energy tensor has this
property and as already have been pointed out by Wainwright (\cite{Wainwrit}),
this model has a power law solution.

Under the action of a similarity group, each physical quantity $\phi$
transforms according to its dimension $q$ under the scale transformation. For
space-times with a transitive similarity group, dimensionless quantities are
therefore spacetime constants. This implies that the ratio of the pressure of
the energy density is constant so that the only possible equation of state is
the usual one in cosmology i.e. $p=\omega\rho$, where $\omega$ is a constant.
In the same way, the existence of homothetic vector implies the existence of
conserved quantities.

In the first place we would like to emphasize that under the hypothesis $K=0,
$ we have found that the space-time $(M,g)$ is self-similar since we have been
able to calculate a homothetic vector field, $X_{H}\in\mathfrak{X}(M),$ that
is to say
\begin{equation}
L_{X}g=2g,
\end{equation}
where
\begin{equation}
X_{H}=\left(  \frac{\int c(t)dt+C_{1}}{c(t)}\right)  \partial_{t}+\left(
1-\left(  \frac{\int c(t)dt+C_{1}}{c(t)}\right)  H\right)  \left(
x\partial_{x}+y\partial_{y}+z\partial_{z}\right)  .
\end{equation}

This kind of space-times have been studied by (Eardly (\cite{22}), Wainwright
(\cite{Wainwrit}), K. Rosquits and R. Jantzen (\cite{Jantzen}) and for reviews
see Carr and Coley (\cite{20}) and Duggal et al (\cite{Duggal})).

As we have mentioned above we are only interested in the case $K=0$ for this
reason eq. (\ref{Lie-general}) yields
\begin{equation}
\rho^{\prime\prime}=\frac{\rho^{\prime2}}{\rho}+\rho^{\prime}\frac{c^{\prime}%
}{c}+A\frac{G}{c^{2}}\rho^{2}.\label{Lie}%
\end{equation}

Now, we apply the standard Lie procedure to this equation. A vector field
\ $X=\xi(t,\rho)\partial_{t}+\eta(t,\rho)\partial_{\rho},$ is a symmetry of
(\ref{Lie}) iff
\begin{align}
\xi_{\rho}+\rho\xi_{\rho\rho}  & =0,\label{sil1}\\
\eta c-2c^{\prime}\rho^{2}\xi_{\rho}-c\rho\eta_{\rho}+c\rho^{2}\eta_{\rho\rho
}-2c\rho^{2}\xi_{t\rho}  & =0,\label{sil2}\\
-cc^{\prime}\rho\xi_{t}-3\rho^{3}AG\xi_{\rho}+\left(  c^{\prime^{2}}%
\rho-cc^{\prime\prime}\rho\right)  \xi+2c^{2}\rho\eta_{t\rho}-c^{2}\rho
\xi_{tt}-2c^{2}\eta_{t}  & =0,\label{sil3}\\
\rho^{2}AG\eta_{\rho}+c^{2}\eta_{tt}-cc^{\prime}\eta_{t}-2\rho AG\eta
-2\rho^{2}AG\xi_{t}+\rho^{2}A\xi\left(  2G\frac{c^{\prime}}{c}-G^{\prime
}\right)   & =0,\label{sil4}%
\end{align}
Solving (\ref{sil1}-\ref{sil4}), we find that
\begin{equation}
\xi=at+b,\qquad\eta=-2a\rho,
\end{equation}
subject to the following constrains, from eq. (\ref{sil3}):
\begin{equation}
c^{\prime\prime}=\frac{c^{\prime2}}{c}-\frac{ac^{\prime}}{at+b}%
,\label{condition1}%
\end{equation}
and from eq. (\ref{sil4})
\begin{equation}
\frac{G^{\prime}}{G}=2\frac{c^{\prime}}{c},\label{condition2}%
\end{equation}
where $a$ and $b$ are numerical constants.

We would like to emphasize that this is the main difference with respect to
our previous work (\cite{to1}) where we found that the model admitted more
symmetries. Note that in this paper we are considering that a $c$-variable
introduces corrections into the curvature tensor, this possibility brings us
to obtain a new eq. (\ref{Lie}) which contains an extra term $\left(
\rho^{\prime}\frac{c^{\prime}}{c}\right)  $ with respect to the employed one
in our previous paper (\cite{to1}).

In this situation we have found that this model only admits two symmetries
$X_{1}=\partial_{t}$ i.e. a movement and $X_{2}=t\partial_{t}-2\rho
\partial_{\rho},$ i.e. a scaling symmetry as it is expected in this kind of
models (self-similar model). In such a way that $\left[  X_{1},X_{2}\right]
=X_{1}$ i.e. they form a $L_{2}$ Lie-algebra, see (\cite{Ibra}).

From (\ref{condition2}) we can see that
\begin{equation}
G\approx c^{2}\qquad i.e.\qquad\frac{G}{c^{2}}=const:=B,
\end{equation}
(where we will assume that $B>0)$ that is to say, that both constants vary but
in such a way that the relationship $\frac{G}{c^{2}}$ remains constant for any
$t$ and independently of the any value of the constants $a$ and $b.$ For this
reason our equation (\ref{Lie}) may be rewritten as follows
\begin{equation}
\rho^{\prime\prime}=\frac{\rho^{\prime2}}{\rho}+\rho^{\prime}\frac{c^{\prime}%
}{c}+A\rho^{2},\label{mercedes}%
\end{equation}
where $A=12\pi\left(  \omega+1\right)  ^{2}B=const>0,\forall\omega,(\omega
\neq-1),$ in such a way that all the restrictions come from eq.
(\ref{condition1}).

The knowledge of one symmetry $X$ might suggest the form of a particular
solution as an invariant of the operator $X$ i.e. as solution of
\begin{equation}
\frac{dt}{\xi}=\frac{d\rho}{\eta},
\end{equation}
this particular solution is known as an invariant solution (generalization of
similarity solution) furthermore an invariant solution is in fact a particular
singular solution.

In order to solve (\ref{mercedes}), we consider the following cases.

\subsubsection{Case I.}

Taking $a=0,b\neq0,$ we get
\begin{equation}
c^{\prime\prime}=\frac{c^{\prime2}}{c},\qquad\Longrightarrow\qquad
c(t)=K_{2}e^{K_{1}t},\label{c_1}%
\end{equation}
where the $\left(  K_{i}\right)  _{i=1}^{2}$ are integration constants. In
this way we find that from eq. (\ref{condition2}) $G$ behaves as:
\begin{equation}
G=K_{2}^{2}e^{2K_{1}t}.
\end{equation}

Therefore the equation (\ref{mercedes}) yields:
\begin{equation}
\rho^{\prime\prime}=\frac{\rho^{\prime2}}{\rho}+\rho^{\prime}K_{1}+A\rho
^{2},\label{nl-c_1}%
\end{equation}
where $X_{1}=\partial_{t}.$ The use of the canonical variables brings us to an
Abel ode:
\begin{equation}
y^{\prime}=-Ax^{2}y^{3}-K_{1}y^{2}-\frac{y}{x},\label{abel_c1_1}%
\end{equation}
where $x=\rho$ and $y=1/\rho^{\prime},$ since we cannot solve it, then we try
to obtain a solution through the invariants

This method brings us to the following relationship
\begin{equation}
\frac{dt}{\xi}=\frac{d\rho}{\eta}\Longrightarrow\rho\approx const:=\rho_{0},
\end{equation}
which seems not to be physical. This particular solution must satisfy eq.
(\ref{mercedes}) which means that $\rho=0.$ With this solution we go back to
eq. (\ref{nf2})
\begin{equation}
3H^{2}=\Lambda c^{2},
\end{equation}
and from eq. (\ref{nf4}) it is obtained the behavior of \textquotedblleft
constant\textquotedblright\ $\Lambda$%
\begin{equation}
\left(  \frac{G^{\prime}}{G}-4\frac{c^{\prime}}{c}\right)  \rho=-\frac
{\Lambda^{\prime}c^{4}}{8\pi G}=0\Longrightarrow\Lambda=\Lambda_{0}=cons.,
\end{equation}
therefore
\begin{equation}
3H^{2}=\Lambda_{0}K_{2}^{2}e^{2K_{1}t}\Longrightarrow f=C_{1}\exp\left(
\frac{\sqrt{\frac{\Lambda_{0}}{3}}}{K_{1}}\exp(K_{1}t)\right)  ,
\end{equation}
with
\begin{equation}
RiemS:=24K_{1}^{4},\qquad RiccS:=36K_{1}^{4}.
\end{equation}

This symmetry has brought us to obtain a super de Sitter solution, with a
energy density vanishing and with a scale factor growing like a power of
exponential functions, while $G$ and $c$ follow an exponential functions that
depends of the constant $K_{1}.$ If for example we fix $K_{1}=1,$ then there
is a sudden singularity in a finite time, in this case both \textquotedblleft
constants\textquotedblright\ $G$ and $c$ are growing functions on time $t.$
While if $K_{1}=-1,$ the scale factor reaches an asymptotic behavior with $G$
and $c$ decreasing on time $t.$ In any case $\Lambda=const.>0,$ i.e. is a true
positive constant.

\subsubsection{Case II.}

Taking $b=0,a\neq0,$ we get that the infinitesimal $X$ is $X_{1}=t\partial
_{t}-2\rho\partial_{\rho}$, which is precisely the generator of the scaling
symmetries. Therefore the invariant solution will be the same than the
obtained one with the dimensional method.

With these values of $a$ and $b$ equation (\ref{condition1}) yields
\begin{equation}
c^{\prime\prime}=\frac{c^{\prime2}}{c}-\frac{c^{\prime}}{t},\qquad
\Longrightarrow\qquad c(t)=K_{2}t^{K_{1}},\label{c_2}%
\end{equation}
as we expected, $c(t)$ follows a power-law solution (self-similar solution),
where $K_{1}$ and $K_{2}$ are numerical constants, $K_{1},K_{2}\neq0$. We
would like to emphasize that with this behavior for $c$ the homothetic vector
field behaves as:%
\begin{equation}
X_{H}=\frac{t}{K_{1}+1}\partial_{t}+\left(  1-\frac{tH}{K_{1}+1}\right)
\left(  x\partial_{x}+y\partial_{y}+z\partial_{z}\right)
\end{equation}
as in the standard model (see \ for example Eardley (\cite{22}) and Wainwrith
(\cite{Wainwrit})).

In this way $G$ behaves as:%
\begin{equation}
G=K_{2}^{2}t^{2K_{1}}.
\end{equation}
Therefore equation (\ref{mercedes}) yields:
\begin{equation}
\rho^{\prime\prime}=\frac{\rho^{\prime2}}{\rho}+\rho^{\prime}\frac{K_{1}}%
{t}+A\rho^{2},\label{nl-c_2}%
\end{equation}
we expect that this ode admits a scaling solution (i.e. that the energy
density follows a power law solution) as it is expected in this kind of models.

The canonical variables bring us to obtain an Abel ode
\begin{equation}
y^{\prime}=x\left(  2+2K_{1}-Ax\right)  y^{3}-(1+K_{1})y^{2}-\frac{y}%
{x},\label{abel_c2_1}%
\end{equation}
where $x=\rho t^{2}$ and $y=\frac{1}{t^{2}\left(  \rho^{\prime}t+2\rho\right)
},$ and which solution is unknown

The solution obtained through invariants is:
\begin{equation}
\frac{dt}{\xi}=\frac{d\rho}{\eta}\Longrightarrow\rho\approx t^{-2},\qquad
\rho=\frac{2(1+K_{1})}{At^{2}},\label{sol_nl-c_2}%
\end{equation}
finding that $K_{1}>-1$ from physical considerations i.e. $\rho>0$ iff
$K_{1}>-1.$ This is the kind of solution expected in a self-similar model (see
\ for example Wainwrith (\cite{Wainwrit}) and Jantzen (\cite{Jantzen})).

From eq. (\ref{nf4}) it is obtained the behavior of \textquotedblleft
constant\textquotedblright\ $\Lambda$%
\begin{equation}
\Lambda^{\prime}=16\pi B\rho\dfrac{c^{\prime}}{c^{3}}\Longrightarrow
\Lambda=-\dfrac{16\pi BK_{1}}{K_{2}^{2}At^{2(1+K_{1})}},
\end{equation}
as it is observed if $K_{1}>0$ then $\Lambda$ is a negative decreasing
function on time $t,$while if $K_{1}<0$ then $\Lambda$ is a decreasing
function on time $t,$ but with the restriction $K_{1}\in\left(  -1,0\right)  .
$ We would like to emphasize that the self-similar relationship
\begin{equation}
\Lambda\thickapprox\dfrac{1}{c^{2}t^{2}}\thickapprox\dfrac{1}{K_{2}%
^{2}t^{2(1+K_{1})}},
\end{equation}
is trivially verified as it is expected in this kind of solutions.

Now, we will calculate the scale factor $f.$ In order to do that we may follow
two ways, in the first one, from eq. (\ref{nf3}) we find that
\begin{equation}
f=f_{0}t^{2/3\left(  \omega+1\right)  },\label{f_cas_2}%
\end{equation}
where $f_{0}=\left(  \dfrac{A_{\omega}A}{2(1+K_{1})}\right)  ^{1/3\left(
\omega+1\right)  },$ observing again that necessarily $K_{1}>-1.$

If we follow our second way it is found that from eq. (\ref{nf2} with $K=0$)%
\begin{equation}
3H^{2}=\dfrac{16\pi B}{At^{2}}\qquad\Longrightarrow\qquad f=f_{0}%
t^{\sqrt{H_{0}}},
\end{equation}
where $H_{0}=\dfrac{16\pi B}{3A},$ but taking into account the value of
$A=12\pi\left(  \omega+1\right)  ^{2}B,$ then $H_{0}=\dfrac{4}{9\left(
\omega+1\right)  ^{2}},$ and hence $\sqrt{H_{0}}=\dfrac{2}{3\left(
\omega+1\right)  },$ as it is expected from eq. (\ref{f_cas_2}). Therefore
this new way does not add any more information.

In this way we found that
\begin{equation}
H=\sqrt{H_{0}}t^{-1},\qquad and\qquad q=\dfrac{d}{dt}\left(  \dfrac{1}%
{H}\right)  -1=\dfrac{1}{\sqrt{H_{0}}}-1.
\end{equation}
therefore $q<0$ iff $\omega\in\left(  -1,-\frac{1}{3}\right)  ,$ note that if
$\omega<-1$ (phantom cosmology)$,$ then $f(t)$ is a decreasing function on
time $t$.

If we make the assumption (scaling symmetry) on the scale factor
\begin{equation}
f\thickapprox ct\qquad\Longrightarrow\qquad f\thickapprox t^{K_{1}+1}%
\end{equation}
in such a way that equating this expression with eq. (\ref{f_cas_2}) then%
\begin{equation}
K_{1}=-\dfrac{1+3\omega}{3\left(  \omega+1\right)  }%
\end{equation}
finding that $K_{1}>0\Longleftrightarrow\omega\in\left(  -1,-\frac{1}%
{3}\right)  .$ Hence, if $K_{1}>0\Longleftrightarrow\omega\in\left(
-1,-\frac{1}{3}\right)  $, $G$ and $c$ are growing functions on time $t,$
while $\Lambda<0,$ i.e. is a negative decreasing function on time $t.$ In
other case, if $K_{1}<0,G$ and $c$ are decreasing functions on time $t,$ while
$\Lambda>0,$ i.e. is a decreasing function on time $t.$

The Kretschmann scalars behave as:%
\begin{equation}
RiemS\thickapprox\frac{1}{K_{2}^{4}t^{4(1+K_{1})}},\qquad\qquad
RiccS\thickapprox\frac{1}{K_{2}^{4}t^{4(1+K_{1})}},
\end{equation}
finding in this way that if $K_{1}<-1$ (forbidden possibility) then both
scalars tend to zero while if $K_{1}>-1,$ then both scalars tend to infinity
i.e. there is a true singularity.

In the first place we would like to emphasize that this is the solution that
we have obtained in our previous works using D.A. (see \cite{to2} and
\cite{to3}). This is due to two reasons. The first one, because D.A. is a
powerful tool that may be used even when the FE are not well outlined. The
second one, because in the previous works we were using only three of the FE,
ignoring eq. (\ref{nf1}) and using eqs. (\ref{nf2},\ref{nf3} and \ref{nf4}).
Therefore the solution found in those papers work well and we refer to them
for all the physical considerations.

In this approach we are supposing that $q<0$ i.e. the universe accelerates due
to a equation of state $\omega\in\left(  -1,-\frac{1}{3}\right)  ,$ but as it
has been pointed out by R.G.Vishwakarma (see \cite{viswa}) the acceleration of
the universe may be explained through different mechanisms in such a way that
$q<0$ while $\omega\in\left[  0,1\right]  .$ In this way we would like to
stress that to solve some of the cosmological problems that present the
standard model, we have found that $K_{1}>-1,$ with $K_{1}=-\dfrac{1+3\omega
}{3\left(  \omega+1\right)  }$ in such a way that if $K_{1}>0 $ iff $\omega
\in\left(  -1,-\frac{1}{3}\right)  $ then $q<0$ and $G$ and $c$ are growing
functions on time $t$, while $\Lambda$ is a \ negative decreasing function on
time $t.$ With $K_{1}<0,$ $q>0$ and $G$ and $c$ are decreasing functions on
time $t$, while $\Lambda$ is a \ positive decreasing function on time $t,$
i.e. there is a narrow relationship between the behavior of $G$ and $c$ and
the sign of $\Lambda$ controlled by $\omega.$

\subsubsection{Case III.}

This case is a generalization of the above case, simply, we will avoid the
singular case. Taking $a,b\neq0$ we get
\begin{equation}
c^{\prime\prime}=\frac{c^{\prime2}}{c}-\frac{ac^{\prime}}{at+b},\quad
\Longrightarrow\quad c(t)=K_{2}\left(  at+b\right)  ^{\frac{K_{1}}{a}%
},\label{c_3}%
\end{equation}
obtaining in this case that $G$ behaves as:%
\begin{equation}
G=K_{2}^{2}\left(  at+b\right)  ^{\frac{2K_{1}}{a}},
\end{equation}
hence equation (\ref{Lie}) yields:
\begin{equation}
\rho^{\prime\prime}=\frac{\rho^{\prime2}}{\rho}+\rho^{\prime}\frac{K_{1}%
}{at+b}+A\rho^{2},\label{nl-c_3}%
\end{equation}
The canonical variables bring us to an Abel ode
\begin{equation}
y^{\prime}=x\left(  2a^{2}+2aK_{1}-Ax\right)  y^{3}-(a+K_{1})y^{2}-\frac{y}%
{x}\label{abel_c3_1}%
\end{equation}
where $x=\rho\left(  at+b\right)  ^{2}$ and $y=\frac{1}{\left(  at+b\right)
^{2}\left(  \rho^{\prime}(at+b)+2a\rho\right)  },$ that is to say, we obtain a
very complicate ode which at this time we do not know how to solve.
Nevertheless, as we have pointed out in (\cite{to2}) the general solution of
this kind of equations are unphysical, for this reason it is sufficient
consider particular solutions obtained through the invariants.

The solution obtained through invariants is:
\begin{equation}
\frac{dt}{\xi}=\frac{d\rho}{\eta}\Longrightarrow\rho=\frac{2a(a+K_{1}%
)}{A\left(  at+b\right)  ^{2}},\label{sol_nl-c_3}%
\end{equation}
where it is observed that $K_{1}>-a$ (but if you choose $a=b=1,$ then we have
the same result as before i.e.~$K_{1}>-1)$. Therefore the scale factor behaves
as:%
\begin{equation}
f=\left(  \dfrac{A_{\omega}A\left(  at+b\right)  ^{2}}{2a(a+K_{1})}\right)
^{1/3\left(  \omega+1\right)  },\qquad\Longrightarrow\qquad f=f_{0}\left(
at+b\right)  ^{2/3\left(  \omega+1\right)  }%
\end{equation}
which is very similar to the last result (see eq. (\ref{f_cas_2}) but in this
occasion this solution is non-singular).

We end finding the behavior of $\Lambda$, as in previous cases taking into
account eq. (\ref{nf4}) it is obtained:%
\begin{equation}
\Lambda^{\prime}=16\pi B\rho\dfrac{c^{\prime}}{c^{3}}\qquad\Longrightarrow
\qquad\Lambda=-\dfrac{16\pi}{AK_{2}^{2}}\dfrac{BaK_{1}}{\left(  at+b\right)
^{2\left(  \frac{K_{1}}{a}+1\right)  }},
\end{equation}
as we can see in this case if $K_{1}>0$ then $\Lambda$ is a negative
decreasing function on time $t.$

The Kretschmann scalars behaves as:%
\begin{equation}
RimS\thickapprox\frac{1}{K_{2}^{4}\left(  at+b\right)  ^{4(1+K_{1})}}%
,\qquad\qquad RiccS\thickapprox\frac{1}{K_{2}^{4}\left(  at+b\right)
^{4(1+K_{1})}},
\end{equation}
showing a non-singular state when $t$ runs to zero.

This solution is very similar to the previous one except that this solution is
non-singular. In fact, the above solution is a particular solution of this one.

\subsection{The non-flat case.}

In this subsection we go next to study the particular case $K\neq0.$ One of
the drawbacks of the above approach is that we need to make the assumption
$K=0$ i.e. our approach is unable of solving the so-called flatness problem.
In order to research if it is possible to solve such problem we go next to
study eq. (\ref{Lie-general}) through the Lie method, seeking symmetries that
allow us to obtain any solution in closed from. But as we will see eq.
(\ref{Lie-general}) does not admit any symmetry in such a way that in order to
obtain a particular solution we will impose a concrete symmetry, but this is
precisely the method that we are trying to avoid, to make assumptions or at
least to make the minor number of assumptions or to make assumptions under any
physical or mathematical (symmetries) well founded reasons. We only explore
one case.

Therefore the equation under study is:%
\begin{equation}
\rho^{\prime\prime}=\frac{\rho^{\prime2}}{\rho}+\rho^{\prime}\frac{c^{\prime}%
}{c}-2\frac{Kc^{2}}{f^{2}}\rho+A\frac{G}{c^{2}}\rho^{2},
\end{equation}
but as
\begin{equation}
divT=0\Longleftrightarrow\rho=A_{\omega}f^{3\left(  \omega+1\right)
},\Longrightarrow f=\left(  \frac{\rho}{A_{\omega}}\right)  ^{\frac
{1}{3\left(  \omega+1\right)  }},\Longrightarrow f=\left(  \rho\right)
^{\frac{1}{3\left(  \omega+1\right)  }},
\end{equation}
and hence
\begin{equation}
\rho^{\prime\prime}=\frac{\rho^{\prime2}}{\rho}+\rho^{\prime}\frac{c^{\prime}%
}{c}-2c^{2}\rho^{a}+A\frac{G}{c^{2}}\rho^{2},\label{carmen}%
\end{equation}
where $a=\frac{3\omega+1}{3\left(  \omega+1\right)  },$ and for simplicity we
have adopted the case $K=1.$

The Lie group method brings us to obtain the following system of pdes%

\begin{align}
\rho{\xi}_{\rho\rho}+{\xi}_{\rho}  & =0,\label{s_32_1}\\
-\rho^{-1}{\eta}_{\rho}-2\frac{c^{\prime}}{c}{\xi}_{\rho}+{\eta}_{\rho\rho
}-2\xi_{t\rho}+\rho^{-2}\eta & =0,\label{s_32_2}\\
\left(  \left(  \frac{c^{\prime}}{c}\right)  ^{2}-\frac{c^{\prime\prime}}%
{c}\right)  \xi+3\rho^{2}\left(  2c^{2}\rho^{a-2}-A\frac{G}{c^{2}}\right)
{\xi_{\rho}}-2\rho^{-1}{\eta_{t}-{\xi_{tt}}+2{\eta_{t\rho}}-}\frac{c^{\prime}%
}{c}{{\xi_{t}}}  & {=}{0,}\label{s_32_3}%
\end{align}%
\[
\eta_{tt}-\frac{c^{\prime}}{c}{\eta_{t}+}\rho^{2}\left(  A\frac{G}{c^{2}%
}-2c^{2}\rho^{a-2}\right)  {\eta_{\rho}}+2\rho^{2}\left(  2c^{2}\rho
^{a-2}-A\frac{G}{c^{2}}\right)  {\xi_{t}}%
\]%
\begin{equation}
+A\rho^{2}\frac{G}{c^{2}}\left(  4\rho^{a-2}\frac{c^{4}}{AG}\frac{c^{\prime}%
}{c}-\frac{G^{\prime}}{G}+2\frac{c^{\prime}}{c}\right)  \xi+2\rho\left(
ac^{2}\rho^{a-2}-A\frac{G}{c^{2}}\right)  \eta=0\label{s_32_4}%
\end{equation}
which has no solution, that is to say, eq. (\ref{carmen}) does not admit any
symmetry, for this reason we will need to follow other approaches.

For example, if we \textquotedblleft impose\textquotedblright\ any particular
symmetry $X,$ maybe we may found some restrictions for the behavior of the
quantities $G,c$ and $\rho.$ We will explore such possibility.

\subsubsection{Case I.}

In this case, we choose $\left(  \xi=1,\eta=0\right)  ,i.e.$ $X=\partial_{t},$
in such a way that from eq.(\ref{s_32_3}) it is obtained the following
restriction%
\begin{equation}
c^{\prime\prime}=\frac{c^{\prime^{2}}}{c}\qquad\Longrightarrow\qquad
c(t)=K_{2}e^{K_{1}t},\label{res1}%
\end{equation}
and from eq.(\ref{s_32_4}) it is obtained the following one%
\begin{equation}
2\left(  2\rho^{a-2}\frac{c^{4}}{AG}+1\right)  \frac{c^{\prime}}{c}%
=\frac{G^{\prime}}{G},\label{res2}%
\end{equation}
hence%
\begin{equation}
2\left(  2\rho^{a-2}\frac{K_{2}^{4}e^{4K_{1}t}}{AG}+1\right)  K_{1}%
=\frac{G^{\prime}}{G},\label{res2_1}%
\end{equation}
as we can see from eq. (\ref{res2}), we cannot obtain the condition $G=Bc^{2}
$ (as in the flat solution) since such condition means that%
\begin{equation}
4\rho^{a-2}\frac{c^{4}}{AG}\frac{c^{\prime}}{c}=\frac{G^{\prime}}{G}%
-2\frac{c^{\prime}}{c}=0\Longleftrightarrow\rho=0,
\end{equation}
that is to say, the energy density vanishes.

In order to find a particular solution to eq. (\ref{res2_1}) we impose the
condition $a=2$ (as mathematical condition) which means that $\omega=-\frac
{5}{3}\ll-1,$ although such possibility is very restrictive (and maybe
unphysical, the ultra phantom equation of state). In this way it is found that%
\begin{equation}
G^{\prime}=2K_{1}G+\frac{4K_{1}K_{2}^{4}e^{4K_{1}t}}{A},\Longrightarrow
G(t)=C_{1}e^{2tK_{1}}+\frac{2}{A}K_{2}^{4}e^{4(tK_{1})},
\end{equation}
where $C_{1}$ is an integration constant, therefore eq. (\ref{carmen}) yields%
\begin{equation}
\rho^{\prime\prime}=\frac{\rho^{\prime2}}{\rho}+\rho^{\prime}K_{1}+\left(
A\frac{C_{1}}{K_{2}^{2}}\right)  \rho^{2},
\end{equation}
since this equation has not an explicit (analytical) solution, note that we
have obtained the same eq. as in section 3.1 case I eq. (\ref{nl-c_1}), we
find again that a particular solution is $\rho=0,$ therefore from the field
equation (\ref{nf2})
\begin{equation}
H^{2}+\frac{K_{2}^{2}e^{2K_{1}t}}{f^{2}}=\frac{\Lambda_{0}}{3}K_{2}%
^{2}e^{2K_{1}t},
\end{equation}
and hence in this case we have found that a solution is:%
\begin{equation}
f=\frac{\sqrt{3}}{\sqrt{\Lambda_{0}}}\left(  \frac{2}{3}+\frac{1}{6}%
\exp\left(  \frac{2\sqrt{3}}{3}\frac{\sqrt{\Lambda_{0}}K_{2}}{K_{1}}\left(
\exp(K_{1}t)+C_{1}\right)  \right)  \exp\left(  \frac{-\sqrt{3}}{3}\frac
{\sqrt{\Lambda_{0}}K_{2}}{K_{1}}\left(  \exp(K_{1}t)-C_{1}\right)  \right)
\right)  ,
\end{equation}
$\allowbreak$since%
\begin{equation}
0=\left(  \frac{G^{\prime}}{G}-4\frac{c^{\prime}}{c}\right)  \rho
=-\frac{\Lambda^{\prime}c^{4}}{8\pi G}\Longrightarrow\Lambda=\Lambda
_{0}=const.>0.
\end{equation}

This solution looks very unphysical or at least very restrictive, a vanishing
energy density and a constant cosmological constant and is very similar to the
obtained one in section 3.1. case I the super de-Sitter solution, with the
same behavior and the same restriction for the numerical constant $K_{1},$
i.e. sudden singularities or asymptotic behavior depending of the sign of
$K_{1}.$ In this case, the employed method does not allow us to obtain the
behavior of $G,$ $c$ and $\Lambda$ since the equation under study does not
admit any symmetry.

\section{Solution II. $divT\neq0.$}

In the previous section we have made the assumption that $div(T)=0$ i.e. the
divergence of the energy--momentum tensor vanishes. Nevertheless we have
obtained as general conservation equation eq. (\ref{def_nf3}) i.e.
\begin{equation}
\rho^{\prime}+3\left(  \omega+1\right)  \rho H=-\left(  \frac{G^{\prime}}%
{G}-4\frac{c^{\prime}}{c}\right)  \rho-\frac{\Lambda^{\prime}c^{4}}{8\pi G},
\end{equation}
i.e. $div(T)=\rho^{\prime}+3\left(  \omega+1\right)  \rho H\neq0$ and we have
assumed a particular case (with perfect mathematical sense) $div(T)=0.$ In
this section we will study the general case $div\left(  T\right)  \neq0.$

The possibility that cosmological and physical considerations may require that
the covariant conservation condition $div(T)=0$ be relaxed has been advanced
by Rastall (\cite{Rastall}), who pointed out that a non-zero divergence of the
energy--momentum tensor has as yet not been ruled out experimentally at all.
In Rastall's theory (\cite{Rastall}), the divergence of $T$ is assumed to be
proportional to the gradient of the scalar curvature $S$, $div(T)=\lambda
grad(S)$ , where
\c{}
$\lambda=constant$, and in fact the modified field equations are equivalent to
standard general relativity with an additional variable $\Lambda$ term. We
refer to the reader to the Harko and Mark work (\cite{Harko1}) to see a matter
creation and thermodynamical approach in this context.

\subsection{The flat case.}

In this case we are going to consider the field equations \textquotedblleft
but\textquotedblright\ without the condition $div(T)=0,$ i.e. the field
equations will be
\begin{align}
2H^{\prime}-2\frac{c^{\prime}}{c}H+3H^{2}  & =-\frac{8\pi G}{c^{2}}p+\Lambda
c^{2},\label{nf1_solII}\\
3H^{2}  & =\frac{8\pi G}{c^{2}}\rho+\Lambda c^{2},\label{nf2_solII}\\
\rho^{\prime}+3\left(  \omega+1\right)  \rho H  & =-\left(  \frac{G^{\prime}%
}{G}-4\frac{c^{\prime}}{c}\right)  \rho-\frac{\Lambda^{\prime}c^{4}}{8\pi
G},\label{nf3_solII}%
\end{align}
but as we can see we have $2$ equations with $5$ unknowns, therefore it is
necessary to make some assumptions, for example in the behavior of $G,c$ and
$\Lambda$ as follows:%
\begin{equation}
G=G_{0}H^{a},\qquad c=c_{0}H^{b},\qquad\Lambda=\Lambda_{0}c^{-2}H^{2},
\end{equation}
where $G_{0},c_{0}$ and $\Lambda_{0}$ are dimensional constants and
$a,b\in\mathbb{R},$ without any restriction i.e. we do not need to assume any
concrete sing or value for these numerical constants. Furthermore, we must
stress that the conditions $K=0,$ together to power law assumptions bring us
to a scaling solution.

Taking into account these assumptions and form eq. (\ref{nf2_solII}) we obtain
$\rho$%
\begin{equation}
\rho=\left(  \dfrac{3-\Lambda_{0}}{d_{0}}\right)  H^{2(1+b)-a},\qquad
i.e.\qquad\rho=\rho_{0}H^{2(1+b)-a},
\end{equation}
where $d_{0}=8\pi G_{0}/c_{0}^{2},$ and taking this relationship into eq.
(\ref{nf3_solII}), we obtain the following ode in quadrature%
\begin{equation}
\alpha\dfrac{H^{\prime}}{H}+3\left(  \omega+1\right)  H=-a\dfrac{H^{\prime}%
}{H}+4b\dfrac{H^{\prime}}{H}-\tilde{K}\dfrac{H^{\prime}}{H},
\end{equation}
and therefore
\begin{equation}
\left(  \alpha+a-4b+\tilde{K}\right)  \dfrac{H^{\prime}}{H^{2}}=-3\left(
\omega+1\right)  ,
\end{equation}
where $\alpha=2(1+b)-a,$ $\tilde{K}=\left(  2(1-b)c_{0}^{2}\Lambda_{0}/8\pi
G_{0}\rho_{0}\right)  ,$ and therefore%
\begin{equation}
\dfrac{H^{\prime}}{H^{2}}=-\dfrac{\left(  \omega+1\right)  \left(
3-\Lambda_{0}\right)  }{2\left(  1-b\right)  },\qquad\Longrightarrow\qquad
H=h_{0}nt^{-1},\qquad\Longrightarrow\qquad f=f_{0}t^{h_{0}n},
\end{equation}
where $n=\dfrac{\left(  \omega+1\right)  \left(  3-\Lambda_{0}\right)
}{2\left(  1-b\right)  },$ and $h_{0}=const>0,$ is an integration constant and
we impose that $b\neq1,\Lambda_{0}\neq3$ and $\omega\neq-1.$ As we can see,
there are some restrictions, for example, if $\omega\in(-1,1]$ then
$b,\Lambda_{0}$ must verify at the same time that $b<1$ and $\Lambda_{0}<3$ or
$b>1$ and $\Lambda_{0}>3.$ Now if $\omega<-1$ then it should exist a
combination between the signs of $b,\Lambda_{0}$ such that $n>0$, in other
way, the radius of the Universe decreases. It is observed that the behavior of
the scale factor does not depend of the constant $a$, i.e. of the behavior of
the gravitational \textquotedblleft constant\textquotedblright.

Once we have obtained the behavior of $f$ i.e. of $H$ we go next to complete
our calculations of the rest of the quantities i.e.%
\begin{equation}
\rho=\rho_{0}H^{2(1+b)-a}\qquad\Longrightarrow\rho=\rho_{0}\left(
h_{0}nt^{-1}\right)  ^{2(1+b)-a}\thickapprox t^{a-2(1+b)},
\end{equation}
with the restrictions%
\begin{equation}
\rho_{0}\left(  h_{0}n\right)  ^{2(1+b)-a}>0,\qquad and\qquad a-2(1+b)<0,
\end{equation}
i.e. we are assuming that the energy density is a positive decreasing function
on time $t$.

It is observed too that if $2b=a,$ then we obtain the particular solution
$\rho\thickapprox t^{-2},$ as well as the realtionship\ $G/c^{2}=const.$ as in
the above cases, i.e. the obtained solution under the assumption $div(T)=0.$
In this way we can se that the $div(T)=0$ case is a particular solution of the
$div(T)\neq0$ case, as one may expected, but we have not any physical or
mathematical (symmetry or integrability condition) reason to assume such relationship.

We end calculating the behavior of the \textquotedblleft
constants\textquotedblright\ $G,c$ and $\Lambda$ i.e.%
\begin{equation}
G=G_{0}h_{0}^{a}n^{a}t^{-a},\qquad c=c_{0}\left(  h_{0}n\right)  ^{b}%
t^{-b},\qquad\Lambda=\Lambda_{0}c_{0}^{-2}\left(  h_{0}n\right)
^{2(b-1)}t^{2(b-1)},
\end{equation}
in such a way that $\Lambda$ will be a decreasing function on time iff $b<1. $

In this way we found that
\begin{equation}
H=h_{0}nt^{-1},\qquad and\qquad q=\dfrac{d}{dt}\left(  \dfrac{1}{H}\right)
-1=\dfrac{1}{h_{0}n}-1=\dfrac{2\left(  1-b\right)  }{h_{0}\left(
\omega+1\right)  \left(  3-\Lambda_{0}\right)  }-1,
\end{equation}
therefore $q<0$ iff $\left\vert 2\left(  1-b\right)  \right\vert <\left\vert
h_{0}\left(  \omega+1\right)  \left(  3-\Lambda_{0}\right)  \right\vert .$

The Kretschmann scalars behave as:%
\begin{equation}
RiemS\thickapprox\frac{1}{c_{0}^{4}t^{4\left(  1-b\right)  }},\qquad\qquad
RiccS\thickapprox\frac{1}{c_{0}^{4}t^{4\left(  1-b\right)  }},
\end{equation}
finding in this way that there is a true singularity.

In this case we have found a scaling solution as it is expected for this kind
of models ($K=0$). As we can see the scale factor $f$ does not depend of $G$
only of $c$ and $\Lambda.$ The energy density $\rho$ depends of $G$ and $c.$
Nevertheless we have not been able to find better restrictions for the
introduced (ad hoc) numerical constants $a$ and $b$ such that they give us
more information about the behavior of the time functions $G$, $c$ and
$\Lambda$

\subsection{The non-flat case.}

As \ in the above section we will study the case $K\neq0$ separately. For this
purpose the FE are now:%
\begin{align}
2H^{\prime}-2\frac{c^{\prime}}{c}H+3H^{2}+\frac{Kc^{2}}{f^{2}}  & =-\frac{8\pi
G}{c^{2}}p+\Lambda c^{2},\label{leticia1}\\
H^{2}+\frac{Kc^{2}}{f^{2}}  & =\frac{8\pi G}{3c^{2}}\rho+\frac{1}{3}\Lambda
c^{2},\label{leticia2}\\
\rho^{\prime}+3\left(  \omega+1\right)  \rho H  & =-\left(  \frac{G^{\prime}%
}{G}-4\frac{c^{\prime}}{c}\right)  \rho-\frac{\Lambda^{\prime}c^{4}}{8\pi
G},\label{leticia3}%
\end{align}
but as we can see we have $2$ equations with $5$ unknowns, therefore it is
necessary to make some assumptions, for example in the behavior of $G,c$ and
$\Lambda$ as follows:%
\begin{equation}
G=G_{0}H^{a},\qquad c=c_{0}H^{b},\qquad\Lambda=\Lambda_{0}c^{-2}H^{2},
\end{equation}
where $G_{0},c_{0}$ and $\Lambda_{0}$ are numerical constants. We must stress
that in this occasion these hypotheses could be unphysical since $K\neq0$ and
we are imposing a scaling behavior typical of the flat case.

Taking into account these assumptions and form eq. (\ref{leticia2}) we obtain
$\rho$%
\begin{equation}
\rho=\left(  \dfrac{3-\Lambda_{0}}{d_{0}}\right)  H^{2(1+b)-a}+\frac
{Kc_{0}H^{2b}}{f^{2}},\qquad i.e.\qquad\rho=\rho_{0}H^{2(1+b)-a}+\frac
{Kc_{0}H^{2b}}{f^{2}},
\end{equation}
where $d_{0}=8\pi G_{0}/c_{0}^{2},$ and taking this relationship into eq.
(\ref{leticia3}), we obtain the following second order for $f$%
\begin{equation}
f^{\prime\prime}=-D_{1}\frac{\left(  f^{\prime}\right)  ^{2b}}{f^{(1+2b)}%
}+D_{2}\dfrac{\left(  f^{\prime}\right)  ^{2}}{f},\label{def_ecu}%
\end{equation}
where%
\begin{equation}
D_{1}=\dfrac{\left(  3\omega+1\right)  }{2}\frac{Kc_{0}^{2}}{\left(
1-b\right)  },\qquad D_{2}=\left(  1-\dfrac{\left(  \omega+1\right)  }%
{2}\dfrac{\left(  3-\Lambda_{0}\right)  }{\left(  1-b\right)  }\right)  .
\end{equation}

In the first place we may note that $\left(  b\neq1\right)  $ and that if
$\omega=-1/3\Longrightarrow D_{1}=0,$ independently of the value of constant
$K$, in this case we are mainly interested in the $K\neq0$ case$.$ If $K=0,$
or $D_{1}=0$ then we obtain again the solution already obtained in the latter
(last) case.

Calculation of Eq. (\ref{def_ecu}). making the following change of variables
it is obtained the first order ode%
\begin{equation}
(x=f,\quad y=\frac{1}{f^{\prime}})\quad\Longrightarrow\quad y^{\prime}%
=D_{1}x^{-(1+2b)}y^{3-2b}-D_{2}x^{-1}y,
\end{equation}
and which solution is:
\begin{equation}
y=\left(  C_{1}x^{2D_{2}(1-b)}+\frac{2(b-1)D_{1}}{\left(  2bD_{2}%
-2D_{2}-2b\right)  }x^{-2b}\right)  ^{\frac{1}{2b-2}},
\end{equation}
therefore%
\begin{equation}
f^{\prime}=\left(  C_{1}f^{2D_{2}(1-b)}+\frac{2(b-1)D_{1}}{\left(
2bD_{2}-2D_{2}-2b\right)  }f^{-2b}\right)  ^{\frac{1}{2(1-b)}},
\end{equation}
and hence%
\begin{equation}
t=\int^{f}\left(  C_{1}u^{2D_{2}(1-b)}+\frac{2(b-1)D_{1}}{\left(
2bD_{2}-2D_{2}-2b\right)  }u^{-2b}\right)  ^{\frac{1}{2(1-b)}}du+C_{2},
\end{equation}

Since we have not obtain information about the behavior of $f$ then we try to
find a particular solution. For this purpose, we observe that eq.
(\ref{def_ecu}) admits the following symmetries:%
\begin{equation}
X_{1}=\partial_{t},\qquad X_{2}=t\partial_{t}+\left(  1-b\right)
f\partial_{f},
\end{equation}
where we would emphasize that $X_{2}$ is a scaling symmetry, maybe induced by
the hypotheses about the behavior of $G,c$ and $\Lambda,$ (scaling relationships).

The invariant solution (particular solution) that induces $X_{2}$ is the
following one:%
\begin{equation}
\dfrac{dt}{t}=\dfrac{df}{\left(  1-b\right)  f}\qquad\Longrightarrow\qquad
f=f_{0}t^{\left(  1-b\right)  },
\end{equation}
with
\begin{equation}
f_{0}=\frac{D_{1}(b-1)}{D_{2}(b-1)-b}\sqrt{\frac{D_{2}(b-1)-b}{D_{1}(b-1)}%
}=\frac{Kc_{0}^{2}\left(  1+3\omega\right)  }{\left(  3\omega+1-\Lambda
_{0}(\omega+1)\right)  }\sqrt{\frac{\left(  3\omega+1-\Lambda_{0}%
(\omega+1)\right)  }{Kc_{0}^{2}\left(  1+3\omega\right)  }}=const.,
\end{equation}
and we necessarily impose that $b<1$ and $b\neq0,$ this means that if
$b\rightarrow1$ them $f\rightarrow f_{0}.$ As we can observe $K\neq0,$ and we
must be careful with the signs since if $K=-1$ then it should be satisfied the
relationship $3\omega+1<\Lambda_{0}(\omega+1).$

In this way we found that
\begin{equation}
H=(1-b)t^{-1},\qquad and\qquad q=\dfrac{d}{dt}\left(  \dfrac{1}{H}\right)
-1=\dfrac{1}{1-b}-1=\dfrac{b}{1-b},
\end{equation}
therefore $q<0$ iff $b<0,$ which implies that $c$ is a growing function on
time $t$.

If we make the assumption (scaling-symmetry) on the scale factor
\begin{equation}
f\thickapprox ct\qquad\Longrightarrow\qquad f\thickapprox t^{\left(
1-b\right)  },
\end{equation}

The Kretschmann scalars behave as:%
\begin{equation}
RiemS\thickapprox\frac{1}{C^{4}t^{4(b-1)}},\qquad\qquad RiccS\thickapprox
\frac{1}{C^{4}t^{4(b-1)}},
\end{equation}
finding in this way that there is a true singularity.

Calculation of $\rho,$ and the possible restrictions for the constants $a$ and
$b.$%
\begin{equation}
\rho=\rho_{0}\left(  1-b\right)  ^{2b-a+2}t^{a-2(b+1)}+K\frac{c_{0}}{f_{0}%
^{2}}\left(  1-b\right)  ^{2b}t^{-2},
\end{equation}
and%
\begin{equation}
G\thickapprox G_{0}t^{-a},\qquad c\thickapprox c_{0}t^{-b},\qquad
\Lambda\thickapprox\Lambda_{0}^{2}t^{2(b-1)}.
\end{equation}

As we have seen, we have only been able to obtain a particular solution.
Imposing so many hypotheses, we lose information and we do not know how to
recover it in such a way that we are not able to know better the behavior of
the time functions $G,c$ and $\Lambda.$

\section{Conclusions.}

In this paper we have studied a perfect fluid FRW model with time-varying
constants \textquotedblleft but\textquotedblright\ taking into account the
possible effects of a $c-$variable into the curvature tensor. In this way, as
other authors have already pointed out, such effects are minimum in the field
equations but they exists and are very restrictive. Under the made hypotheses,
we have seen that the Einstein tensor has covariant divergence zero, in this
way we have imposed that the right hand of the field equations has a vanishing
divergence too i.e.
\begin{equation}
div\left(  \frac{8\pi G}{c^{4}}T+g\Lambda\right)  =0.\label{paula-m}%
\end{equation}

In this way we have obtained the set of the new FE, emphasizing the fact that
we can recover eq. (\ref{paula-m}) from the field equations as in the standard
case i.e. deriving one of them and simplifying with the other one.

In other to solve the resulting FE we have considered the following cases. In
the first case we have imposed the condition $div(T)=0,$ as a particular case
of eq. (\ref{paula-m}) and we have studied the flat and non-flat cases. We
have needed to make such distinction because with the employed method we are
not able of solving the so called flatness problem, for this reason we have
needed to study separately then.

The flat case under this hypothesis is the already studied one in our previous
works (see \cite{to1}) and under this new considerations i.e. taking into
account the effects of a $c$-variable into the curvature tensor, we have shown
that the scaling solutions the only one while in (\cite{to1}), without this
new assumption, we obtained more solutions, i.e. we obtained other solutions
apart from the scaling solution. We would like to emphasize that it has been
obtained as integration condition that the \textquotedblleft
constants\textquotedblright\ $G$ and $c$ must verify the relationship
$G/c^{2}=cosnt.$ in spite of the fact that both \textquotedblleft
constants\textquotedblright\ vary. This result is in agreement with our
scaling solution obtained in our previous work (see \cite{to1}) but in those
works we needed to make such relationship as assumption.

We have also shown that this model is self-similar since we have been able of
obtaining a non-trivial homothetic vector field. This result is agreement with
the obtained scaling solution as it is well known.

With the obtained solution, if we want to solve the acceleration of the
universe i.e. $q<0$, then \ the time function $G$ and $c$ are growing function
on time $t$, while $\Lambda$ is a negative decreasing function on time $t,$ in
this case the equation of state belong to the interval: $\omega\in\left(
-1,-1/3\right)  $. In other cases $G$ and $c$ are decreasing function while
$\Lambda$ is a positive decreasing function and in this case the equation of
state belongs to the interval: $\omega\in\left[  -1/3,1\right]  .$

The non-flat case does not admit any symmetry and the particular studied case
looks very unphysical.

The second class of studied models verifies the condition $div(T)\neq0,$ i.e.
without imposing any restriction to eq. (\ref{paula-m}) and we have studied
again the flat and non-flat cases. To solve these cases we have needed to make
scaling assumptions on the behavior of the time functions $G$, $c$ and
$\Lambda$. These assumptions work well in the flat case (self-similar case)
but in the non-flat case seem very restrictive. In the flat case, it is found
for the made assumptions, that there is a relationship between the numerical
constants that determine the behavior of $G$ and $c$ that brings us to obtain
again the particular case $div(T)=0,$ i.e. such case could be seen as a
particular solution of the more general case $div(T)\neq0.$

\begin{acknowledgments}
We wish to acknowledge to Javier Aceves his translation into English of this paper.
\end{acknowledgments}

\end{document}